# Prediction of Giant Tunneling Magnetoresistance in RuO$_2$/TiO$_2$/RuO$_2$ (110) Antiferromagnetic Tunnel Junctions

Yuan-Yuan Jiang,[1,2] Zi-An Wang,[1,2] Kartik Samanta,[3] Shu-Hui Zhang,[4] Rui-Chun Xiao,[5] W. J. Lu,[1] Y. P. Sun,[6,1,7]
Evgeny Y. Tsymbal,[3,*] and Ding-Fu Shao[1,†]

[1] *Key Laboratory of Materials Physics, Institute of Solid State Physics, HFIPS, Chinese Academy of Sciences, Hefei 230031, China*

[2] *University of Science and Technology of China, Hefei 230026, China*

[3] *Department of Physics and Astronomy & Nebraska Center for Materials and Nanoscience,*
*University of Nebraska, Lincoln, Nebraska 68588-0299, USA*

[4] *College of Mathematics and Physics, Beijing University of Chemical Technology, Beijing 100029, People's Republic of China*

[5] *Institute of Physical Science and Information Technology, Anhui University, Hefei 230601, China*

[6] *High Magnetic Field Laboratory, HFIPS, Chinese Academy of Sciences, Hefei 230031, China*

[7] *Collaborative Innovation Center of Microstructures, Nanjing University, Nanjing 210093, China*

Using first-principles quantum-transport calculations, we investigate spin-dependent electronic and transport properties of antiferromagnetic tunnel junctions (AFMTJs) that consist of (110)-oriented antiferromagnetic (AFM) metal RuO$_2$ electrodes and an insulating TiO$_2$ tunneling barrier. We predict the emergence of a giant tunneling magnetoresistance (TMR) effect in a wide energy window, a series of barrier layer thicknesses, and different interface terminations, indicating the robustness of this effect. We show that the predicted TMR cannot be explained in terms of the global transport spin-polarization of RuO$_2$ (110) but is well understood based on matching the momentum-dependent spin-polarized conduction channels of the two RuO$_2$ (110) electrodes. We predict oscillations of TMR with increasing barrier thickness, indicating a non-negligible contribution from the perfectly epitaxial interfaces. Our work helps the understanding of the physics of TMR in AFMTJs and aids in realizing efficient AFM spintronic devices.

## I. INTRODUCTION

Spintronics utilizes a spin degree of freedom and magnetic order parameters as state variables to encode information [1]. The electrical read-out of information in spintronic devices requires a strong transport response to the variation of the magnetic order parameters. A typical and widely used spintronic device is a magnetic tunnel junction (MTJ) that is composed of two ferromagnetic (FM) metal electrodes separated by a non-magnetic insulating tunneling barrier [2-7]. In MTJs, low and high resistance states occur for parallel and antiparallel magnetization of the two electrodes, respectively. This effect, known as tunneling magnetoresistance (TMR), offers an ON/OFF ratio as high as a few hundred percent, sufficient for accurate read-out. Due to TMR, MTJs can serve as building blocks of magnetic random-access memories (MRAMs) for data storage and processing [8].

The TMR effect in MTJs has been widely understood in terms of a spin-polarized tunneling current that is controlled by the relative magnetization orientation of the two FM electrodes. This mechanism is often empirically quantified by Julliere's formula, $TMR = \frac{2p_1 p_2}{1-p_1 p_2}$, where $p_i$ ($i = 1, 2$) is the transport spin polarization of the *i*-th FM electrode in a MTJ [2]. Based on this formula, a larger spin polarization of the electrodes favors a larger TMR. While Julliere's formula offers a qualitative explanation of TMR in MTJs, in crystalline MTJs where the transverse wave vector is conserved in the tunneling process, a more accurate description should take into account symmetry matching of the incoming and outcoming Bloch states in the electrodes and evanescent states in the barrier [9]. In particular, matching of the majority-spin $\Delta_1$ band in the Fe (001) electrode to the $\Delta_1$ evanescent state in the MgO (001) barrier layer is responsible for a large positive spin polarization and giant values of TMR predicted for crystalline Fe/MgO/Fe (001) MTJs [10, 11]. These concepts seem to rule out using antiferromagnetic (AFM) metals as electrodes in MTJs, due to the spin degeneracy and hence $p_i = 0$ expected for antiferromagnets with their zero net magnetization.

This understanding has been challenged by the recent theoretical [12-15] and experimental [16, 17] demonstrations of a sizable TMR effect in AFM tunnel junctions (AFMTJs). AFMTJs represent tunnel junctions with two AFM electrodes, where the TMR effect occurs in response to a change of the relative orientation of the AFM order parameters in the electrodes, known as the Néel vectors. The TMR effect relies on the conservation of the transverse momentum in the process of tunneling that requires epitaxial AFMTJs with a well-defined crystalline texture propagating across the whole junction. The possibility of TMR in AFMTJs opens perspectives for employing antiferromagnets in MRAMs, making use of their advantages of being robust against magnetic perturbations, not producing stray fields, and exhibiting ultrafast spin dynamics [18, 19].

In our previous work [12], using first-principles quantum transport calculations, we have explored $RuO_2/TiO_2/RuO_2$ (001) AFMTJs. These AFMTJs employed $RuO_2$ – high-temperature AFM metal exhibiting a spin-split band structure [20, 21]. We predicted that TMR in these AFMTJs was controlled by the matching of the spin-polarized conduction channels in the two $RuO_2$ (001) electrodes. As a result, a large TMR appeared in these AFMTJs, even in the presence of the globally spin-neutral currents. This implies that the net spin polarization of the electrodes is not essential for obtaining a large TMR in AFMTJs.

Recent theoretical predictions also show that the antiferromagnets with spin-split band structures, including the noncollinear antiferromagnets [22, 23] and certain types of collinear antiferromagnets [24-28], dubbed altermagnets [29, 30], are capable of supporting longitudinal spin-polarized currents [31]. For example, it has been predicted that the spin currents occur along the crystallographic directions different from (001) in $RuO_2$ [28]. Specifically, the presence of a spin-polarized longitudinal current with a large polarization $p_i$ is expected to directly support TMR in the relevant AFMTJs, according to the conventional Julliere's picture. It would be interesting to explore the role of this contribution to the total TMR response associated with the spin-dependent Fermi surface of $RuO_2$.

To address this question, we consider an AFMTJ based on $RuO_2$ electrodes that are stacked in the (110) plane, where the crystallographic [110] direction of $RuO_2$ supports spin-polarized transport providing a "direct" contribution to TMR similar to a conventional MTJ. We perform first-principles quantum-transport calculations of TMR in $RuO_2/TiO_2/RuO_2$ (110) AFMTJs and find a giant effect for a series of $TiO_2$ barrier thicknesses. We argue that the predicted TMR effect cannot be explained by the conventional picture based on the globally spin-polarized current emitted by the $RuO_2$ (110) electrode by rather originates from the matching of the spin-polarized conduction channels in the two $RuO_2$ (110) electrodes. These results uncover the important physics of the TMR effect which may be useful for the practical realization of AFMTJs.

## II. THEORETICAL METHODS

First-principles calculations are performed based on density functional theory (DFT) [32] as implemented in the Vienna *ab initio* simulation package (VASP) [33, 34]. The pseudopotentials are described using the projector augmented wave (PAW) method [35], and the exchange-correlation functional is treated within the generalized gradient approximation (GGA) developed by Perdew-Burke-Ernzerhof (PBE) [36]. In the calculations, the cutoff energy for the plane-wave expansion is set to 500 eV, and the k-point grid is set to $16 \times 16 \times 16$ to sample the irreducible Brillouin zone. The GGA+U [37, 38] method with $U_{eff}$ = 2 eV on Ru 4*d* orbitals and $U_{eff}$ = 5 eV on Ti 3*d* orbitals is employed in the calculations for $RuO_2$ and $TiO_2$. The Fermi surfaces are calculated using the Wannier90 code [39] with the maximally localized Wannier functions [40, 41] and visualized by FermiSurfer [42].

Transport properties are calculated using the nonequilibrium Green's function formalism (DFT+NEGF approach) [43, 44], as implemented in QuantumATK, Synopsys QuantumATK [45] using the atomic structures relaxed by VASP. In QuantumATK, we set the cut-off energy of 100 Ry and use the nonrelativistic SG15 pseudopotentials [46], and k-point meshes of $12 \times 12 \times 12$ for bulk $RuO_2$ and $TiO_2$ and $11 \times 11 \times 101$ for $RuO_2/TiO_2/RuO_2$ (110) AFMTJ. The spin-polarized GGA+U [37, 38] method with $U_{eff}$ = 1.2 eV on Ru 4*d* orbitals and $U_{eff}$ = 5 eV on Ti 3*d* orbitals is used in the calculations. These parameters have been well tested to ensure that the electronic structure around $E_F$ calculated by QuantumATK is consistent with that calculated by VASP. Transmission functions are calculated using k-point meshes of $401 \times 401$ in the two-dimensional (2D) Brillouin zone of $RuO_2$ (110) and $RuO_2$ (110) based AFMTJs.

## III. RESULTS AND DISCUSSION

The AFM metal $RuO_2$ [20] has a rutile structure with two spin sublattices $Ru_A$ and $Ru_B$ (Fig. 1(a,b)). Its Néel vector is pointing along the [001] direction, and the Néel temperature is reported to be above 300 K [20]. $RuO_2$ can be considered as a C-type antiferromagnet with strong *intra*-sublattice coupling along the [001] direction. As a result, a globally spin-neutral current along this direction is carried by the staggered Néel spin currents on the two magnetic sublattices, resulting in a giant TMR effect [12] and a field-like spin-transfer torque (STT) that enables deterministic switching of the $RuO_2$ (001) Néel vector [47]. The magnetic space group of $RuO_2$ is $P4_2'/mnm'$. It supports fully compensated antiferromagnetism with a spin-split electronic band structure [21]. Figure 1(c) shows the calculated band structure of $RuO_2$, indicating a pronounced spin splitting along the high-symmetry Γ-M and Z-A directions. Figure 1(d) displays the associated spin-up and spin-down Fermi surfaces of $RuO_2$. They can be transformed to each other by a 90° rotation around the [001] direction. Such momentum-dependent spin splitting is responsible for various spin-dependent transport properties [31, 48-52].

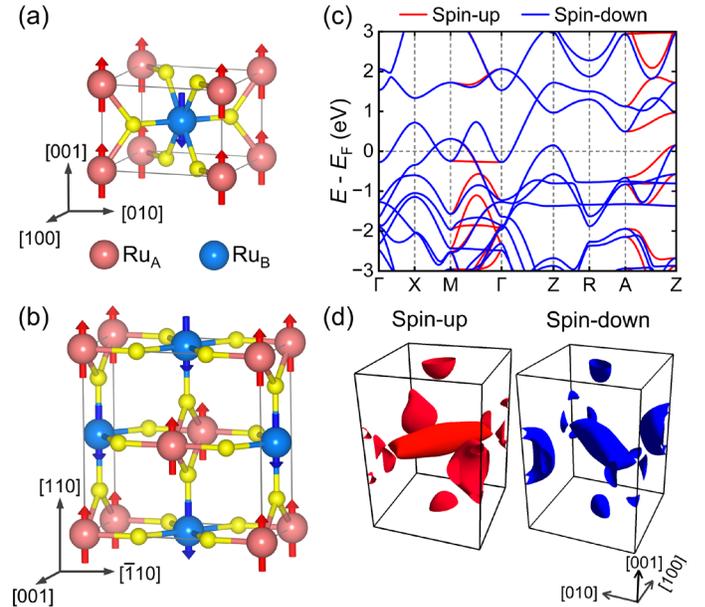

**FIG. 1: (a)** The atomic and magnetic structures of a $RuO_2$ unit cell. **(b)** The atomic and magnetic structures of a $RuO_2$ supercell stacked in the (110) plane. **(c)** The band structure of $RuO_2$. **(d)** The spin-up and spin-down Fermi surfaces of $RuO_2$.

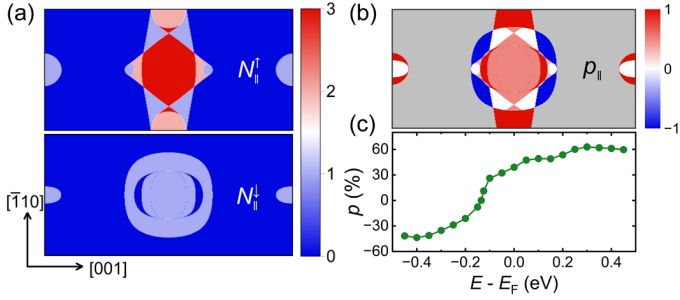

**FIG. 2:** (a) The conduction channels in the 2D Brillouin zone of $RuO_2$ (110) for spin up ($N_\parallel^\uparrow$) and spin down ($N_\parallel^\downarrow$) at the Fermi energy. (b) The spin polarization of conduction channels ($p_\parallel$) at the Fermi energy for $RuO_2$ (110), where the gray color indicates regions with $N_\parallel^\uparrow = N_\parallel^\downarrow = 0$. (c) The global transport spin polarization of $RuO_2$ (110) as a function of energy.

The spin-split Fermi surface of $RuO_2$ supports a longitudinal spin-polarized currents along the [110] or [$\bar{1}$10] directions [28]. This is evident from the calculated ballistic transmission of bulk $RuO_2$ along the [110] direction which reflects the number of conduction channels, i.e., the number of the propagating Bloch states in $RuO_2$ [110]. In the calculation, we used a supercell of $RuO_2$ along the [110] direction shown in Figure 1(b). Figure 2(a) displays the number conduction channels $N_\parallel^\uparrow$ ($N_\parallel^\downarrow$) contributed by the spin-up (spin-down) Fermi surface at different transverse wave vector $\vec{k}_\parallel$ in the 2D Brillouin zone of $RuO_2$ (110) [56]. We find a region of a finite $N_\parallel^\uparrow$ around the zone center, resulting in the maximum $N_\parallel^\uparrow = 3$. There are also some small pockets of $N_\parallel^\uparrow = 1$ at the left and right edges of the zone. The regions of a finite $N_\parallel^\downarrow$ have smaller area located around the zone center and at the left and right edges of the zone. We find that $N_\parallel^\downarrow = 1$ in all these regions of the spin-down Fermi surface.

Figure 2(b) shows the $\vec{k}_\parallel$-dependent spin polarization that is defined as follows:

$$p_\parallel(\vec{k}_\parallel) = \frac{N_\parallel^\uparrow - N_\parallel^\downarrow}{N_\parallel^\uparrow + N_\parallel^\downarrow}. \quad (1)$$

The fully spin-polarization ($p_\parallel = \pm 100\%$) appears in the regions of a finite $N_\parallel^{\uparrow,\downarrow}$ and no overlap between spin-up and spin-down conduction channels. The region around zone center exhibits a relatively small spin polarization ($p_\parallel = 50\%$ in the pink color area and $p_\parallel = 0$ in the white color area in Fig. 2(b)).

Figure 2(c) shows the calculated total transport spin polarization as a function of energy that is defined as follows:

$$p = \frac{\sum_{\vec{k}_\parallel} N_\parallel^\uparrow - N_\parallel^\downarrow}{\sum_{\vec{k}_\parallel} N_\parallel^\uparrow + N_\parallel^\downarrow}. \quad (2)$$

As expected (and in contrast to $RuO_2$ (001)), the total spin polarization is non-zero despite the antiferromagnetism of $RuO_2$. We find that $p = 39\%$ at the Fermi energy ($E_F$), which is comparable to the spin polarization of representative ferromagnetic metals such as Fe, Co, and Ni [53-55]. The spin polarization is enhanced with the increase of energy, reaching a maximum value of $p = 63\%$ at $E = 0.3$ eV, reduced at lower energy, and changes sign at around $E = -0.135$ eV.

Next, we consider a $RuO_2/TiO_2/RuO_2$ (110) AFMTJ where $RuO_2$ (110) serves as electrodes and $TiO_2$ (110) (Fig. 3(a)) as the barrier material. Due to the same rutile structure and similar lattice constants of bulk $RuO_2$ and $TiO_2$, such epitaxial AFMTJ is viable in practice. To characterize the evanescent states in bulk $TiO_2$ (110), we calculate the complex band structure at the Γ point (Fig. 3(b)) and the lowest decay rates of the evanescent states in TiO$_2$ (110) at the Fermi energy (Fig. 3(c)). We find that TiO$_2$ (110) exhibits the smallest decay rates around the vertical midline of the 2D Brillouin zone. Figure 4(a) shows the atomic structure of a RuO$_2$/TiO$_2$/RuO$_2$ (110) heterostructure, including 2 TiO$_2$ monolayers in the center and 6 RuO$_2$ monolayers on each side. As seen from the layer-resolved density of states (DOS) in Figure 4(b), the Fermi energy lies in the middle of the bandgap of TiO$_2$, while some nonvanishing local DOS in the bandgap of the TiO$_2$ layer is due to the metal-induced gap states resulting from RuO$_2$ electrodes. As shown in Supplemental Material [56], with the increasing TiO$_2$ layer thickness, the bandgap opens wider thus sustaining the insulating character of TiO$_2$ and the tunneling transport regime in the AFMTJ.

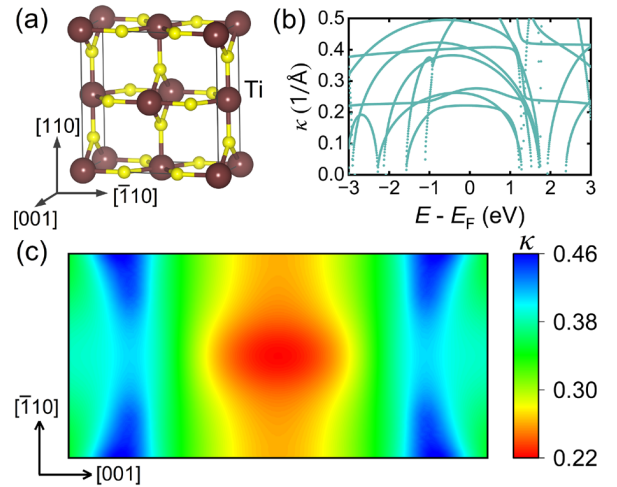

**FIG. 3:** (a) The atomic structures of a $TiO_2$ supercell stacked in the (110) plane. (b) The complex band structure of $TiO_2$ (110) at the Γ point. (c) The lowest decay rates of the evanescent states in $TiO_2$ (110) as a function of $\vec{k}_\parallel$ at the Fermi energy.

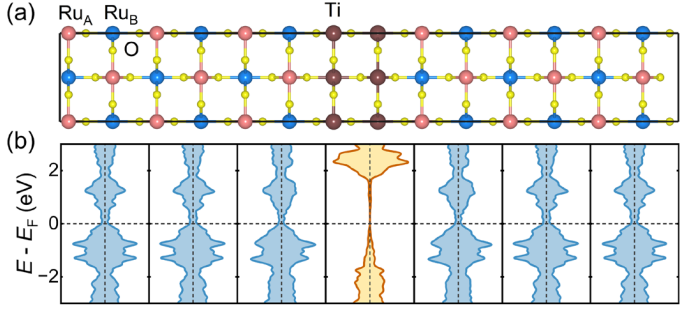

**FIG. 4:** **(a, b)** The atomic structure **(a)** and layer-resolved density of states (DOS) **(b)** of the RuO$_2$/TiO$_2$/RuO$_2$ (110) heterostructure. Each DOS panel contains two $MO_2$ monolayers ($M$ = Ru, Ti).

The RuO$_2$/TiO$_2$/RuO$_2$ (110) heterostructure shown in Figure 4(a) serves as the scattering region in an AFMTJ by connecting this region to two semi-infinite RuO$_2$ (110) electrodes. This geometry allows calculating electron transmission, as described in Sec. II. Figure 5(a,b) shows the calculated $\vec{k}_\parallel$-resolved transmission for the parallel (P) state of the AFMTJ, $T_P^\sigma(\vec{k}_\parallel)$, and for the antiparallel (AP) state, $T_{AP}^\sigma(\vec{k}_\parallel)$, where the $\sigma = \uparrow$ or $\downarrow$ is the spin index. We find that only conduction channels around the vertical midline in the 2D Brillouin zone contribute to the transmission. This is due to the evanescent states in TiO$_2$ having the lowest decay rates in this region (Fig. 3(c)) and thus being mostly supportive to electron transmission. For the P state, the distribution of $T_P^\uparrow(\vec{k}_\parallel)$ qualitatively reflects that of $N_\parallel^\uparrow$ (compare top panels in Figs. 2(a) and 5(a)), while $T_P^\downarrow(\vec{k}_\parallel)$ is significantly suppressed (Fig. 5 (a), bottom panel), indicating that spin-up electrons dominate in the tunneling process. On the other hand, for the AP state, $T_{AP}^\sigma(\vec{k}_\parallel)$ vanishes at those $\vec{k}_\parallel$ where $|p_\parallel| = 100\%$ and is finite only at $\vec{k}_\parallel$ where $p_\parallel$ is small with the maximum of $T_{AP}^\sigma(\vec{k}_\parallel)$ appearing at $\vec{k}_\parallel$ with $p_\parallel = 0$ (compare Figs. 2(b) and 5(b)). These facts indicate the $T_P^\sigma(\vec{k}_\parallel)$ and $T_{AP}^\sigma(\vec{k}_\parallel)$ are largely controlled by the matching of the spin polarization $p_\parallel$ of the conduction channels of the two electrodes.

Figure 5(c) shows the total transmission as a function of energy for the RuO$_2$/TiO$_2$/RuO$_2$ (110) AFMTJ in the P state ($T_P$) and AP state ($T_{AP}$). The $T_P$ is always greater than $T_{AP}$, leading to a positive TMR ratio ($(T_P - T_{AP})/T_{AP}$) (Fig. 5(d)). We find notable $TMR = 200\%$ at $E_F$, much larger than that predicted by Julliere's formula $TMR = \frac{2p_1 p_2}{1 - p_1 p_2}$, using $p_1 = p_2 = 39\%$. Moreover, the maximum TMR at energies above $E_F$ does not appear at $E = 0.3$ eV where $p_1$ and $p_2$ rich a maximum, and the TMR around $E = -0.135$ eV is still very large even though $p_1 = p_2 = 0$. These facts indicate that TMR cannot be described in terms of the total transport spin polarization, but requires knowledge of its distribution in the momentum space.

We note here that while we are using the concept of $\vec{k}_\parallel$-dependent spin polarization $p_\parallel(\vec{k}_\parallel)$ to qualitatively analyze TMR in terms of the Fermi surface matching, this quantity can be used for the *quantitative* prediction of TMR. A more rigorous description requires using the interface transmission function and its spin polarization [57, 58]. The latter takes into account not only the momentum- and spin-dependent Bloch states in the electrode, but also the evanescent states in the barrier as well as the transmission across the interface. The purpose of the present analysis is therefore just to emphasize the deficiency of the total spin polarization and the importance of the Fermi surface

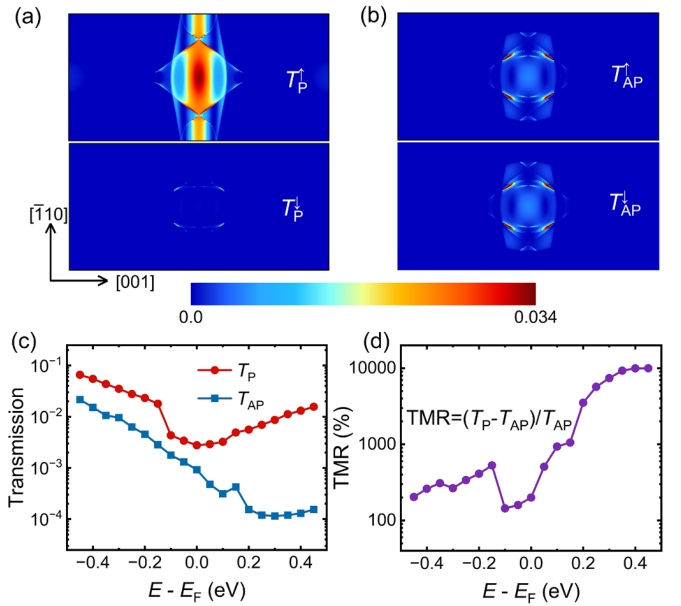

**FIG. 5:** The calculated $\vec{k}_\parallel$-resolved transmission in the 2D Brillouin zone for the AFMTJ in parallel (P) **(a)** and antiparallel (AP) **(b)** states. **(c)** The total transmission as a function of energy for the AFMTJ in the P state (red dots) and the AP state (blue dots). **(d)** TMR as a function of energy.

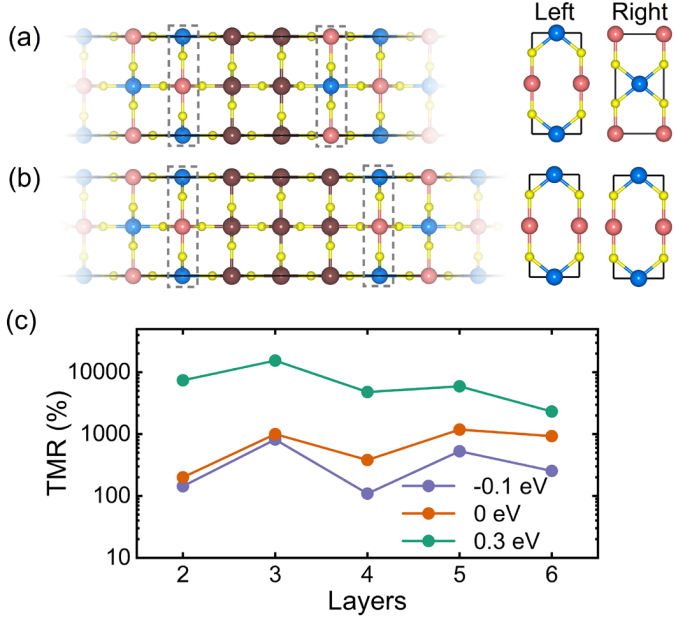

**FIG. 6:** The barrier and interface atomic structure for RuO$_2$|TiO$_2$|RuO$_2$ (110) AFMTJs with TiO$_2$ thickness of 2 layers **(a)** and 3 layers **(b)**. **(c)** TMR as a function of the number of TiO$_2$ monolayers for different energies.

matching in the qualitative picture of TMR in AFMTJs based on spin-split antiferromagnets.

Next, we calculate TMR in RuO$_2$/TiO$_2$/RuO$_2$ (110) AFMTJs with larger thickness of the TiO$_2$ barrier. In the ideal crystalline AFMTJs, changing TiO$_2$ thickness alters the interface structure, as shown in Figures 6 (a,b). For all barrier thicknesses considered, we find a positive TMR. This indicates that the TMR mainly originates from the spin-polarized conduction channels of bulk RuO$_2$ (110). However, we find that TMR in AFMTJs with an odd number of TiO$_2$ monolayers is always larger than TMR in AFMTJs with an even number of TiO$_2$ monolayers, leading to an oscillation of TMR as a function of barrier thickness (Fig. 6(c)). We attribute this phenomenon to the effect of interface. As shown in Figure 6(a), for an AFMTJ with an even number of TiO$_2$ monolayers, the left and right interfacial RuO$_2$ monolayers are asymmetric and can be transformed to each other by a half-unit cell translation along the vertical direction. In this case, the P (AP) state of the AFMTJ has AP (P) interfacial magnetic moments in the horizontal Ru-chains in the two electrodes. This makes the bulk Néel vectors of the electrodes aligned oppositely to the alignment of the interfacial moments. As a result, the large transmission of the AFMTJ for the P-aligned Néel vectors is reduced by the AP-aligned interfacial moments, while the low transmission of the AFMTJ for the AP-aligned Néel vectors is enhanced by the P-aligned interfacial moments, thus reducing the overall TMR. On the contrary, the interfacial RuO$_2$ layers are the same for an AFMTJ with an odd number of TiO$_2$ monolayers. As a result, the P (AP) state of the AFMTJ has P (AP) interfacial moments in the two electrodes. This matching enhances TMR. As evident from Figure 6(c)), the oscillatory TMR appears at different energies. It is slightly suppressed at high energy where TMR is large, while it is more pronounced at $E = -0.1$ eV where TMR has a minimum. These oscillations reflect the competition of bulk and interfacial contributions to TMR.

In order to further understand the influence of the interface structure on TMR, we replace the TiO$_2$ barrier in the AFMTJ with a vacuum layer of ~5 Å to construct a RuO$_2$|□|RuO$_2$ (110) AFMTJ (□ denotes the vacuum layer). We fix the left electrode and shift the right electrode to obtain different interface configurations, as shown in Figure 7(a). The two interfaces in configuration (1) correspond to these in the RuO$_2$/TiO$_2$/RuO$_2$ (110) AFMTJ with an even number of TiO$_2$ monolayers, while the interfaces in configuration (2) represent these in the RuO$_2$/TiO$_2$/RuO$_2$ (110) with an odd number of TiO$_2$ monolayers. Configuration (3) is obtained by applying a quarter-unit-cell translation of the right electrode along the in-plane diagonal direction. Configuration (4) is obtained from configuration (3) by applying a half-unit-cell translation along the vertical direction in the right electrode. As seen from Fig. 7(b), TMR is significantly larger for configuration (2) than for configuration (1), which is expected due to the enhancement of TMR for configuration (2) and its reduction for configuration (1) by the interfacial magnetic moments [56]. On the other hand, a moderate TMR of the same magnitude is calculated for configurations (3) and (4). This is due to the misaligned horizontal Ru chains at the two interfaces, which averages out the interfacial effect on TMR [56].

Interface effects in AFMTJs are important due to interface roughness and disorder being inevitable in experimental conditions. In this regard, the previous predictions of large magnetoresistive effects in AFM spin valves [59-61] and AFMTJs [62-64] are not expected to be robust to interface roughness and disorder [65, 66]. These predictions employed spin-degenerate AFM metals where the bulk contribution to TMR could not occur. The predicted large effects entirely relied on perfect interfaces and switching the interfacial magnetic moments between parallel and antiparallel like in conventional MTJs. In contrast, $RuO_2$ exhibits spin-dependent band structure which is largely responsible for TMR in $RuO_2/TiO_2/RuO_2$ (110) AFMTJs. While the interface magnetic structure contributes to TMR, its contribution is not dominant and therefore even in the presence of interface roughness the predicted large TMR effects are expected to survive. On the other hand, modern film-growth techniques are capable of fabricating high-quality epitaxial heterostructures with the atomic scale control of the interface structure. Using these techniques, it may be possible to manufacture $RuO_2/TiO_2/RuO_2$ (110) AFMTJs with a well-controlled interface structure and observe TMR oscillations predicted in this work.

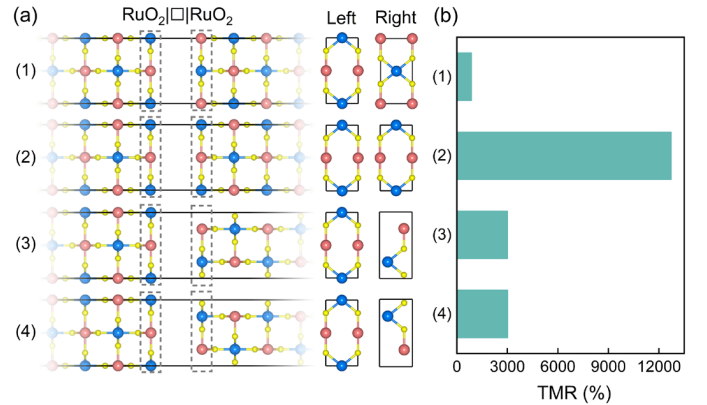

**FIG. 7:** (a) The supercells and interface atomic structures for RuO$_2$|□|RuO$_2$ (110) AFMTJs. (b) The calculated TMR for the four AFMTJs in Fig. (a) at the Fermi energy.

The giant TMR effect in $RuO_2$ (110) based AFMTJs implies a possibility of a large STT in these junctions. However, since the longitudinal current in $RuO_2$ (110) is globally spin-polarized, the generated STT is expected to be mostly damping-like [67-69]. This kind of STT is capable to drive an ultrafast oscillation of the Néel vector [70], but is not able to realize its deterministic switching. An accurate and efficient write-in may be realized by applying an in-plane current in the $RuO_2$ (110) free layer along the [001] direction. Such a current is globally spin-neutral, but staggered, i.e. represents a Néel spin current [47]. For an AFMTJ with a nanoscale width and asymmetric boundary conditions, the Néel spin current can generate a net field-like STT for the deterministic switching of the $RuO_2$ free-layer.

Furthermore, since antiferromagnets such as $RuO_2$ (110) host an unbalanced $p_{\parallel}(\vec{k}_{\parallel})$, they could serve as counter electrodes in conventional MTJs with a single FM electrode, where the matching of the unbalanced $p_{\parallel}(\vec{k}_{\parallel})$ in AFM and FM electrodes generates TMR [71-73]. It is interesting both from the fundamental point of view and from the practical perspective, as it allows using a spin-polarized current from an AFM electrode to generate STT on an FM electrode for magnetization switching and eliminates the pinning layer required in conventional MTJs.

## IV. SUMMARY

In summary, we have investigated spin-polarized transport properties of $RuO_2/TiO_2/RuO_2$ (110) AFMTJs using first-principles quantum-transport calculations. We predicted a giant TMR effect in these junctions and showed that it is robust to the change of electron energy, barrier layer thickness, and the interface termination. The predicted TMR effect cannot be explained by the conventional picture based on the globally spin-polarized current produced by the $RuO_2$ (110) antiferromagnet by rather originates from the matching of the spin-polarized conduction channels in the two $RuO_2$ (110) electrodes. We found TMR oscillations with the increase of $TiO_2$ barrier thickness that reflect a non-negligible contribution from the perfectly epitaxial interfaces. Our work helps the understanding of the physics of TMR and aids in realizing functional AFMTJs in practice.

## ACKNOWLEDGMENTS

This work was supported by the National Key Research and Development Program of China (Grant Nos. 2021YFA1600200, 2022YFA1403203), the National Science Foundation of China (NSFC Grants No. 12241405, 12274411, 52250418, 12174019, 12274412, 12204009, U2032215), the Basic Research Program of the Chinese Academy of Sciences Based on Major Scientific Infrastructures (No. JZHKYPT-2021-08), and the CAS Project for Young Scientists in Basic Research (No. YSBR-084) Research at University of Nebraska-Lincoln was supported by the Division of Materials Research of the National Science Foundation (NSF grant No. DMR-2316665). The calculations were performed at Hefei Advanced Computing Center.

* tsymbal@unl.edu
† dfshao@issp.ac.cn

[1] E. Y. Tsymbal and I. Žutić, Eds., *Spintronics Handbook: Spin Transport and Magnetism*, 2-nd edition (CRC press, 2019).
[2] M. Julliere, Tunneling between ferromagnetic films, *Phys. Lett. A* **54**, 225 (1975).
[3] R. Meservey and P. M. Tedrow, Spin-polarized electron tunneling, *Phys. Rep.* **238**, 173 (1994).
[4] J. S. Moodera, L. R. Kinder, T. M. Wong, and R. Meservey, Large magnetoresistance at room temperature in ferromagnetic thin film tunnel junctions, *Phys. Rev. Lett.* **74**, 3273 (1995).
[5] E. Y. Tsymbal, O. N. Mryasov, and P. R. LeClair, Spin-dependent tunneling in magnetic tunnel junctions, *J. Physics: Condensed Matter* **15**, R109-R142 (2003).
[6] S. Yuasa, T. Nagahama, A. Fukushima, Y. Suzuki, and K. Ando, Giant room-temperature magnetoresistance in single-crystal Fe/MgO/Fe magnetic tunnel junctions, *Nat. Mater.* **3**, 868 (2004).
[7] S. Parkin, C. Kaiser, A. Panchula, P. M. Rice, B. Hughes, M. Samant, and S. H. Yang, Giant tunnelling magnetoresistance at room temperature with MgO (100) tunnel barriers, *Nat. Mater.* **3**, 862 (2004).
[8] A. V. Khvalkovskiy, D. Apalkov, S. Watts, R. Chepulskii, R. S. Beach, A. Ong, X. Tang, A. Driskill-Smith, W. H. Butler, P. B.Visscher, D. Lottis, E. Chen, V. Nikitin, and M. Krounbi, Basic principles of STT-MRAM cell operation in memory arrays, *J. Phys. D* **46**, 074001 (2013).

[9] Ph. Mavropoulos, N. Papanikolaou, and P. H. Dederichs, Complex band structure and tunneling through ferromagnet/insulator/ferromagnet junctions, *Phys. Rev. Lett.* **85**, 1088 (2000).

[10] W. H. Butler, X. G. Zhang, T. C. Schulthess, and J. M. MacLaren, Spin-dependent tunneling conductance of Fe|MgO|Fe sandwiches, *Phys. Rev. B* **63**, 054416 (2001).

[11] J. Mathon and A. Umerski, Theory of tunneling magnetoresistance of an epitaxial Fe/MgO/Fe (001) junction, *Phys. Rev. B* **63**, 220403 (2001).

[12] D.-F. Shao, S. H. Zhang, M. Li, C. B. Eom, and E. Y Tsymbal, Spin-neutral currents for spintronics, *Nat. Commun.* **12**, 7061 (2021).

[13] L. Šmejkal, A. Birk Hellenes, R. González-Hernández, J. Sinova, and T. Jungwirth, Giant and tunneling magnetoresistance in unconventional collinear antiferromagnets with nonrelativistic spin-momentum coupling, *Phys. Rev. X* **12**, 011028 (2022).

[14] J. Dong, X. Li, G. Gurung, M. Zhu, P. Zhang, F. Zheng, E. Y. Tsymbal, and J. Zhang, Tunneling magnetoresistance in noncollinear antiferromagnetic tunnel junctions, *Phys. Rev. Lett.* **128**, 197201 (2022).

[15] G. Gurung, D.-F. Shao, and E. Y. Tsymbal, Extraordinary Tunneling Magnetoresistance in Antiferromagnetic Tunnel Junctions with Antiperovskite Electrodes, arXiv:2306.03026 (2023).

[16] P. Qin, H. Yan, X. Wang, H. Chen, Z. Meng, J. Dong, M. Zhu, J. Cai, Z. Feng, X. Zhou, L. Liu, T. Zhang, Z. Zeng, J. Zhang, C. Jiang, and Z. Liu, Room-temperature magnetoresistance in an all-antiferromagnetic tunnel junction, *Nature* **613**, 485 (2023).

[17] X. Chen, T. Higo, K. Tanaka, T. Nomoto, H. Tsai, H. Idzuchi, M. Shiga, S. Sakamoto, R. Ando, H. Kosaki, T. Matsuo, D. Nishio-Hamane, R. Arita, S. Miwa, and S. Nakatsuji, Octupole-driven magnetoresistance in an antiferromagnetic tunnel junction, *Nature* **613**, 490 (2023).

[18] V. Baltz, A. Manchon, M. Tsoi, T. Moriyama, T. Ono, and Y. Tserkovnyak, Antiferromagnetic spintronics, *Rev. Mod. Phys.* **90**, 015005 (2018).

[19] T. Jungwirth, X. Marti, P. Wadley, and J. Wunderlich, Antiferromagnetic spintronics, *Nat. Nanotechnol.* **11**, 231 (2016).

[20] T. Berlijn, P. C. Snijders, O. Delaire et al., Itinerant Antiferromagnetism in $RuO_2$, *Phys. Rev. Lett.* **118**, 077201 (2017).

[21] K.-H. Ahn, A. Hariki, K.-W. Lee, and J. Kuneš, $RuO_2$ Antiferromagnetism in as d-wave Pomeranchuk instability, *Phys. Rev. B* **99**, 184432 (2019).

[22] J. Železný, Y. Zhang, C. Felser, and B. Yan, Spin-polarized current in noncollinear antiferromagnets, *Phys. Rev. Lett.* **119**, 187204 (2017).

[23] G. Gurung, D.-F. Shao, and E. Y. Tsymbal, Transport Spin Polarization of Noncollinear Antiferromagnetic Antiperovskites, *Phys. Rev. Mater.* **5**, 124411 (2021).

[24] S. Hayami, Y. Yanagi, and H. Kusunose, Momentum-dependent spin splitting by collinear antiferromagnetic ordering. *J. Phys. Soc. Jpn*. **88**, 123702 (2019).

[25] L.-D. Yuan, Z. Wang, J.-W. Luo, E. I. Rashba, and A. Zunger, Giant momentum-dependent spin splitting in centrosymmetric low-Z antiferromagnets. *Phys. Rev. B* **102**, 014422 (2020).

[26] L.-D. Yuan, Z. Wang, J.-W. Luo, and A. Zunger, A. Prediction of low-Z collinear and noncollinear antiferromagnetic compounds having momentum-dependent spin splitting even without spin-orbit coupling. *Phys. Rev. Mater.* **5**, 014409 (2021).

[27] M. Naka, S. Hayami, H. Kusunose, Y. Yanagi, Y. Motome, and H. Seo, Spin current generation in organic antiferromagnets, *Nat. Commun.* **10**, 4305 (2019).

[28] R. González-Hernández, L. Šmejkal, K. Výborný, Y. Yahagi, J. Sinova, T. Jungwirth, J. Železný, Efficient electrical spin-splitter based on non-relativistic collinear antiferromagnetism, *Phys. Rev. Lett.* **126**, 127701 (2021).

[29] L. Šmejkal, J. Sinova, and T. Jungwirth, Beyond Conventional Ferromagnetism and Antiferromagnetism: A Phase with Nonrelativistic Spin and Crystal Rotation Symmetry, *Phys. Rev. X* **12**, 031042 (2022).

[30] L. Šmejkal, J. Sinova, and T. Jungwirth, Emerging ResearchLandscape of Altermagnetism, *Phys. Rev. X* **12**, 040501 (2022).

[31] H. Bai, L. Han, X. Y. Feng, Y. J. Zhou, R. X. Su, Q. Wang, L. Y. Liao, W. X. Zhu, X. Z. Chen, F. Pan, X. L. Fan, and C. Song, Observation of Spin Splitting Torque in a Collinear Antiferromagnet $RuO_2$, Phys. Rev. Lett. **128**, 197202 (2022).

[32] P. Hohenberg and W. Kohn, Inhomogeneous Electron Gas, *Phys. Rev.* **136**, B864 (1964).

[33] G. Kresse and J. Hafner, Ab initio molecular dynamics for liquid metals, *Phys. Rev. B* **47**, 558 (1993).

[34] G. Kresse and J. Furthmuller, Efficient iterative schemes for ab initio total-energy calculations using a plane-wave basis set, *Phys. Rev. B* **54**, 11169 (1996).

[35] G. Kresse and D. Joubert, From ultrasoft pseudopotentials to the projector augmented-wave method, *Phys. Rev. B* **59**, 1758 (1999).

[36] J. P. Perdew, K. Burke, and M. Ernzerhof, Generalized Gradient Approximation Made Simple, *Phys. Rev. Lett.* **77**, 3865 (1996).

[37] V. V. Anisimov, J. Zaanen, and O. K. Andersen, Band theory and Mott insulators: Hubbard U instead of Stoner I, *Phys. Rev. B* **44**, 943 (1991).

[38] S. L. Dudarev, G. A. Botton, S. Y. Savrasov, C. J. Humphreys, and A. P. Sutton, Electron-energy-loss spectra and the structural stability of nickel oxide: An LSDA+U study, *Phys. Rev. B* **57**, 1505 (1998).

[39] A. A. Mostofi, J. R. Yates, G. Pizzi, Y. S. Lee, I. Souza, D. Vanderbilt, and N. Marzari, An updated version of wannier90: A tool for obtaining maximally-localised Wannier functions, *Comput. Phys. Commun*. **185**, 2309 (2014).

[40] N. Marzari and D. Vanderbilt, Maximally localized generalized Wannier functions for composite energy bands, *Phys. Rev. B* **56**, 12847 (1997).

[41] I. Souza, N. Marzari, and D. Vanderbilt, Maximally localized Wannier functions for entangled energy bands, *Phys. Rev. B* **65** (2001).

[42] M. Kawamura, FermiSurfer: Fermi-surface viewer providing multiple representation schemes, *Comput. Phys. Commun*. **239**, 197 (2019).

[43] J. Taylor, H. Guo, and J. Wang, Ab initio modeling of quantum transport properties of molecular electronic devices, *Phys. Rev. B* **63**, 245407 (2001).

[44] M. Brandbyge, J.-L. Mozos, P. Ordejón, J. Taylor, and K. Stokbro, Density-functional method for nonequilibrium electron transport, *Phys. Rev. B* **65**, 165401 (2002).

[45] S. Smidstrup, T. Markussen, P. Vancraeyveld et al., QuantumATK: an integrated platform of electronic and atomic-scale modelling tools, *J. Phys. Condens. Matter* **32**, 015901 (2020).

[46] D. R. Hamann, Optimized norm-conserving Vanderbilt pseudopotentials, *Phys. Rev. B* **88**, 085117 (2013).

[47] D. F. Shao, Y. Y. Jiang, J. Ding, S. H. Zhang, Z. A. Wang, R. C. Xiao, G. Gurung, W. J. Lu, Y. P. Sun, and E. Y. Tsymbal, Neel Spin Currents in Antiferromagnets, *Phys. Rev. Lett.* **130**, 216702 (2023).

[48] A. Bose, N. J. Schreiber, R. Jain, D. -F. Shao, H. P. Nair, J. Sun, X. S. Zhang, D. A. Muller, E. Y. Tsymbal, D. G. Schlom, and D. C. Ralph, Tilted spin current generated by the collinear antiferromagnet ruthenium dioxide, *Nat. Electron.* **5**, 267 (2022).

[49] D. -F. Shao, S.-H. Zhang, R.-C. Xiao, Z.-A. Wang, W. J. Lu, Y. P. Sun, and E. Y. Tsymbal, Spin-neutral tunneling anomalous Hall effect, *Phys. Rev. B* **106**, L180404 (2022).

[50] S. Karube, T. Tanaka, D. Sugawara, N. Kadoguchi, M. Kohda, and J. Nitta, Observation of Spin-Splitter Torque in Collinear Antiferromagnetic $RuO_2$, *Phys. Rev. Lett.* **129**, 137201 (2022).

[51] L. Šmejkal, R. González-Hernández, T. Jungwirth, and J. Sinova, Crystal Hall effect in collinear antiferromagnets, *Sci. Adv.* **6**, eaaz8809 (2020).

[52] Z. Feng, X. Zhou, L. Šmejkal, L. Wu, Z. Zhu, H. Guo, R. González-Hernández, X. Wang, H. Yan, P. Qin, X. Zhang, H. Wu, H. Chen, C. Jiang, M. Coey, J. Sinova, T. Jungwirth, and Z. Liu, An anomalous Hall effect in altermagnetic ruthenium dioxide, *Nat. Electron.* **5**, 735 (2022).

[53] R. J. Soulen Jr., J. M. Byers, M. S. Osofsky, B. Nadgorny, T. Ambrose, S. F. Cheng, P. R. Broussard, C. T. Tanaka, J. Nowak, J. S. Moodera, A. Barry, and J. M. D. Coey, Measuring the spin polarization of a metal with a superconducting point contact. *Science* **282**, 85 (1998).

[54] S. K. Upadhyay, A. Palanisami, R. N. Louie, and R. A. Buhrman, Probing ferromagnets with Andreev reflection. *Phys. Rev. Lett.* **81**, 3247 (1998).

[55] I. I. Mazin, How to Define and Calculate the Degree of Spin Polarization in Ferromagnets. *Phys. Rev. Lett.* **83**, 1427 (1999).

[56] See supplemental material for the electronic structure of bulk $RuO_2$ (110), the density of states (DOS) of the $RuO_2/TiO_2/RuO_2$ (110) AFMTJ, the transmission in $RuO_2/TiO_2/RuO_2$ (110) AFMTJ with different $TiO_2$ thickness, and the Transmission in $RuO_2/\square/RuO_2$ (110) AFMTJ with different atomic configurations.

[57] K. D. Belashchenko, E. Y. Tsymbal, M. van Schilfgaarde, D. Stewart, I. I. Oleinik, and S. S. Jaswal, Effect of interface bonding on spin-dependent tunneling from the oxidized Co surface, *Phys. Rev. B* **69**, 174408 (2004).

[58] K. D. Belashchenko and E. Y. Tsymbal, Tunneling magnetoresistance: Theory, in *Spintronics Handbook: Spin Transport and Magnetism*, 2nd Edition, edited by E. Y. Tsymbal and I. Žutić, (CRC Press, Boca Raton, FL, 2019), p. 525.

[59] A. S. Núñez, R. A. Duine, P. Haney, and A. H. MacDonald, Theory of spin torques and giant magnetoresistance in antiferromagnetic metals, *Phys. Rev. B* **73**, 214426 (2006).

[60] P. M. Haney, D. Waldron, R. A. Duine, A. S. Núñez, H. Guo, and A. H. MacDonald, Ab initio giant magnetoresistance and current-induced torques in Cr/Au/Cr multilayers, *Phys. Rev. B* **75**, 174428 (2007).

[61] Y. Xu, S. Wang, and K. Xia, Spin-transfer torques in antiferromagnetic metals from first principles, *Phys. Rev. Lett.* **100**, 226602 (2008).

[62] P. Merodio, A. Kalitsov, H. Ba, V. Baltz, and M. Chshiev, Spin-dependent transport in antiferromagnetic tunnel junctions, *Appl. Phys. Lett.* **105**, 122403 (2014).

[63] M. Stamenova, R. Mohebbi, J. Seyed-Yazdi, I. Rungger, and S. Sanvito, First-principles spin-transfer torque in CuMnA|GaP|CuMnAs junctions, *Phys. Rev. B* **95**, 060403 (2017).

[64] H. Saidaoui, A. Manchon, and X. Waintal, Robust spin transfer torque in antiferromagnetic tunnel junctions, *Phys. Rev. B* **95**, 134424 (2017).

[65] H. B. M. Saidaoui, A. Manchon, and X. Waintal, Spin transfer torque in antiferromagnetic spin valves: From clean to disordered regimes, *Phys. Rev. B* **89**, 174430 (2014).

[66] A. Manchon, Spin diffusion and torques in disordered antiferromagnets, *J. Phys. Condens. Matter* **29**, 104002 (2017).

[67] O. V. Gomonay and V. M. Loktev, Spin transfer and current-induced switching in antiferromagnets, *Phys. Rev. B* **81**, 144427 (2010).

[68] R. Cheng, M. W. Daniels, J.-G. Zhu, and D. Xiao, Ultrafast switching of antiferromagnets via spin-transfer torque, *Phys. Rev. B* **91**, 064423 (2015).

[69] O. Gomonay, V. Baltz, A. Brataas, and Y. Tserkovnyak, Antiferromagnetic spin textures and dynamics, *Nat. Phys.* **14**, 213 (2018).

[70] R. Cheng, D. Xiao, and A. Brataas, Terahertz Antiferromagnetic Spin Hall Nano-Oscillator, *Phys. Rev. Lett.* **116**, 207603 (2016).

[71] K. Samanta, D.-F. Shao, T. R. Paudel, and E. Y. Tsymbal, Giant tunneling magnetoresistance in magnetic tunnel junctions with a single ferromagnetic electrode, *Abstracts of the 68-th Annual MMM Conference* (Dallas, Texas, 2023).

[72] K. Samanta, Y.-Y. Jiang, T. R. Paudel, D.-F. Shao, and E. Y. Tsymbal, Tunneling magnetoresistance in magnetic tunnel junctions with a single ferromagnetic electrode, *arXiv*:2310.02139 (2023).

[73] B. Chi, L. Jiang, Y. Zhu, G. Yu, C. Wan, J. Zhang, and X. Han, Altermagnetic tunnel junctions of $RuO_2/TiO_2/CrO_2$, *arXiv*: 2309.09561 (2023).